# An electron hole doping and soft x-ray spectroscopy study on La$_{1-x}$Sr$_x$Fe$_{0.75}$Ni$_{0.25}$O$_{3-\delta}$

Selma Erat[1,2], Artur Braun[1*], Alejandro Ovalle[1], Cinthia Piamonteze[3],

Zhi Liu[4], Thomas Graule[1,5], Ludwig J. Gauckler[2]

[1]*Laboratory for High Performance Ceramics, Empa- Swiss Federal Laboratories for Materials & Testing Research, CH- 8600 Dübendorf, Switzerland*

[2]*Department of Nonmetallic Inorganic Materials, ETH Zürich- Swiss Federal Institute of Technology, CH- 8053 Zürich, Switzerland*

[3]*Swiss Light Source, Paul Scherrer Institute, 5232 Villigen, Switzerland*

[4] *Advanced Light Source, Lawrence Berkeley National Laboratory, Berkeley CA 94720, USA*

[5]*Technische Universität Bergakademie Freiberg, D -09596 Freiberg, Germany*

---

[*] Author to whom correspondence should be submitted

**ABSTRACT**

The conductivity of the electron hole and polaron conductor $La_{1-x}Sr_xFe_{0.75}Ni_{0.25}O_{3-\delta}$, a potential cathode material for intermediate temperature solid oxide fuel cells, was studied for $0 \leq x \leq 1$ and for temperatures $300\ K \leq T \leq 1250\ K$. In LaSrFe-oxide, an $ABO_3$ type perovskite, A-site substitution of the trivalent $La^{3+}$ by the divalent $Sr^{2+}$ causes oxidation of $Fe^{3+}$ towards $Fe^{4+}$, which forms conducting electron holes. Here we have in addition a B-site substitution by Ni. The compound for $x = 0.5$ is identified as the one with the highest conductivity ($\sigma \sim 678$ S/cm) and lowest activation energy for polaron conductivity ($E_p = 39$ meV). The evolution of the electronic structure was monitored by soft x-ray Fe and oxygen K-edge spectroscopy. Homogeneous trend for the oxidation state of the Fe was observed. The variation of the ambient temperature conductivity and activation energy with relative Sr content (x) shows a correlation with the ratio of $(e_g/e_g+t_{2g})$ in Fe L3 edge up to x=0.5. The hole doping process is reflected by an almost linear trend by the variation of the pre-peaks of the oxygen K-edge soft x-ray absorption spectra.

# INTRODUCTION

Cathodes of batteries and solid oxide fuel cells (SOFC) are typically based on transition metal (TM) oxides. Often the TM ions have a non-integer average oxidation state, frequently referred to as mixed valence. The well known example $Fe_3O_4$ has two $Fe^{3+}$ and one $Fe^{2+}$ ions per formula unit. The present work is about LaSrFeNi-oxide, a perovskite with $ABO_3$ structure. Adding large rare earth (RE) cations to the TM oxides imposes a metrical situation on the structure which facilitates formation of phases with particular properties like metal insulator transitions or charge ordering, and even spectacular phenomena like superconductivity and colossal magnetoresistance. These TM oxides have a higher oxidation state and their ground state is dominated by O(2p) band holes and not by the metal 3d electrons.

In the parent compound $LaFeO_3$, iron is in the high spin, trivalent insulating oxidation state ($3d^5$), making it a charge transfer insulator. Substitution of the $La^{3+}$ on the A-site by $Sr^{2+}$ causes an oxidation of the iron towards $Fe^{4+}$ with $3d^5\underline{L}$ state, $\underline{L}$ denoting an electron ligand hole from the oxygen. This hole acts as an effective charge carrier. The temperature dependence of the electric conductivity in this class of materials follows often an exponential and is known as polaron activated conductivity. With increasing temperature, the TM is getting reduced due to oxygen loss, and the conductivity characteristic changes from polaron - semiconductor – like to metallic with a ~ 1/T decrease.

The maximum conductivity of LaSrFe-oxide is around 500°C, a temperature range relevant for intermediate temperature SOFC. So far, $La_{0.5}Sr_{0.5}FeO_3$ has the highest conductivity in the LSF series, but substitution on the B-site with nickel increases the conductivity significantly to around 800 S/cm. Chiba et al. have carried out an extensive structure and conductivity study on $La_{1-x}Sr_xFe_{1-y}Ni_yO_3$ [1] and found a clear correlation between conductivity and crystallographic structure.

Dho and Hur also find that Ni doped $LaM1_{0.5}M2_{0.5}O_3$ (M1, 2 denote Mn, Fe, V, Cr, Co, Ni) has always the highest conductivity in that series [2]. Similar holds for the $Sr_3FeMO_7$ (M denotes Fe, co, Ni) Ruddlesdon Popper perovskite, where the conductivity increases from Fe to Ni [3], while the value of the p-d transfer integral $T_\sigma$ decreases from Fe to Ni.

The hole doping process by A-site substitution in LSF-oxide has been well studied [4-6] with x-ray spectroscopy. A detailed study on the B site substitution was carried out by Kumar et al. [7]. In analogy to the A-site substitution study by Abbate et al. [8] in $LaFeO_3$, they find that substitution of Sr in the system creates new spectral intensity features in the oxygen NEXAFS spectra, along with increasing conductivity.

In the present work, we extend the previous studies towards LSF-Ni and present high temperature conductivity data, ambient temperature structure data and electronic structure proved with atomic multiplet calculation. Possible explanation for transport properties which is caused by d- and p-type holes monitored by absorption spectroscopy is made in detail.

**EXPERIMENTAL**

The $La_{1-x}Sr_xFe_{0.75}Ni_{0.25}O_{3-\delta}$ (LSFN) with x=0.0, 0.25, 0.50, 0.75, 1.0 was prepared by solid state reaction. The precursors $La_2O_3$ (>99.99 %), $SrCO_3$ (99.9 %), $Fe_2O_3$ (>99.0 %) and NiO (99.8 %) were mixed in stoichiometric proportions, calcined at 1200°C for 4 h and then sintered at 1400°C for 12 h. The oxygen deficiency δ was obtained by thermogravimetry. X-ray powder diffractograms (XRD) were collected with a Philips X'Pert PRO-MPD diffractometer at ambient temperature, operated at 40kV and 40 mA, with Cu-K$_\alpha$ (λ=1.5405 Å) radiation. The diffractograms were measured in the step width of 0.02° in the angular range of 20° ≤ 2θ ≤ 80°. Rietveld structure refinement was performed with GSAS [9, 10].

For the conductivity measurements, the calcined powders were pressed into bars with dimensions of about 5 mm x 3 mm x 25 mm and sintered at 1400°C for 12h. The $SrFe_{0.75}Ni_{0.25}O_3$ is melt upon sintering. Therefore, no more conclusions about the conductivity and electronic structure of this sample will made. Four Pt terminals were made on the sintered bars using Pt paste (CL11-5100, W. C. Heraeus GmbH & Co. KG, Germany) and calcined with a heating/cooling rate 5 K/min up to 1000°C with 45 min. dwelling time at 1000°C, and then cooled to room temperature. These Pt terminals were attached with Pt-wires around the bars, approximately 10 mm distant from each other. The resistivities of the samples were measured by using four point probe technique using a milli-Ohmmeter RESISTOMAT 2318 (Burster Praezisionmesstechnik GmbH & Co. KG, Germany).

X-ray absorption (XAS) spectra at ambient temperature were recorded at the Advanced Light Source in Berkeley at beamline 9.3.2, the end station of which has an operating energy range of 200-1200 eV and an energy resolution of 1/10000. The vacuum chamber base pressure was around 5x10-10 Torr or lower. Signal detection was made in the sample target current mode. Powder samples were dispersed on conducting carbon tape and then mounted on a copper sample holder. Iron L-edge spectra were recorded from 690 to 750 eV. Oxygen K-edge spectra were recorded from 520 to 560 eV in steps of 0.1 eV.

**RESULTS and DISCUSSIONS**

**Crystallographic structure**

The X-ray diffraction patterns of the samples in the series of the LSFN samples are shown in Figure 1. Without any closer inspection, the patterns would basically suggest a cubic symmetry. The visually best match with a cubic structure comes for $SrFe_{0.75}Ni_{0.25}O_3$, followed by $LaFe_{0.75}Ni_{0.25}O_3$. The other three diffractograms show minor deviations from cubic symmetry. Structure refinement by Rietveld analysis [11] however shows that the sample with $LaFe_{0.75}Ni_{0.25}O_3$ actually has an orthorhombic structure with space group Pbnm (62). The samples with x=0.25, 0.50 and 0.75 also exhibit rhombohedral structure with minor contamination by a tetragonal phase, possibly due to the presence of $La_{1.71}Sr_{0.19}NiO_{3.9}$ (reference pattern JCPDS 01-081-2084) phase. $SrFe_{0.75}Ni_{0.25}O_3$ shows two sets of reflections, both phases of which were identified with cubic phases with space group Pm-3m. Inclusion of a second phase was justified by the distinct splitting of reflections. The two unit cell parameters for the two estimated phases are 3.8490(1) Å, 3.8554(0) Å, with a relative abundance of 82% and 18%, respectively. The refined structural parameters for the samples are listed in Table 1.

**Table 1**: Refined structure parameters for $La_{1-x}Sr_xFe_{0.75}Ni_{0.25}O_{3-\delta}$.

|  | a(Å) | b(Å) | c(Å) | V(Å$^3$) |
|---|---|---|---|---|
| x=0.00 | 5.5051(5) | 7.8304(6) | 5.5303(5) | 238.40(3) |
| x=0.25 | 5.4662 | 5.4662 | 13.5268 | 350.02 |
| x=0.50 | 5.4564(8) | 5.4564(8) | 13.421(5) | 346.07(8) |
| x=0.75 | 5.4478 | 13.3400 | 13.3400 | 342.882 |
| x=1.00 | 3.8490(1) | 3.8490(1) | 3.8490(1) | 57.025(5) |
|  | 3.8554(0) | 3.8554(0) | 3.8554(0) | 57.311(4) |

Atomic positions for Pbnm are 4(c) (0, 0, 1/4) for La; 4(b) (0, 0, 1/4) for Fe/Ni and (4c) (1/2, 1/4, 0) for O1 and (8d) (1/4, 1/2, 1/4) for O2. Atomic positions for R-3c are 6(a) (0, 0, 1/4) for La; 6(b) (0, 0, 0) for Fe/Ni and 18(e) (1/2, 0, 1/4) for O. Atomic positions for Pm-3m are 6(a) (1/2, 1/2, 1/2) for Sr; 6(b) (0, 0, 0) for Fe/Ni and 18(e) (1/2, 0, 0) for O.

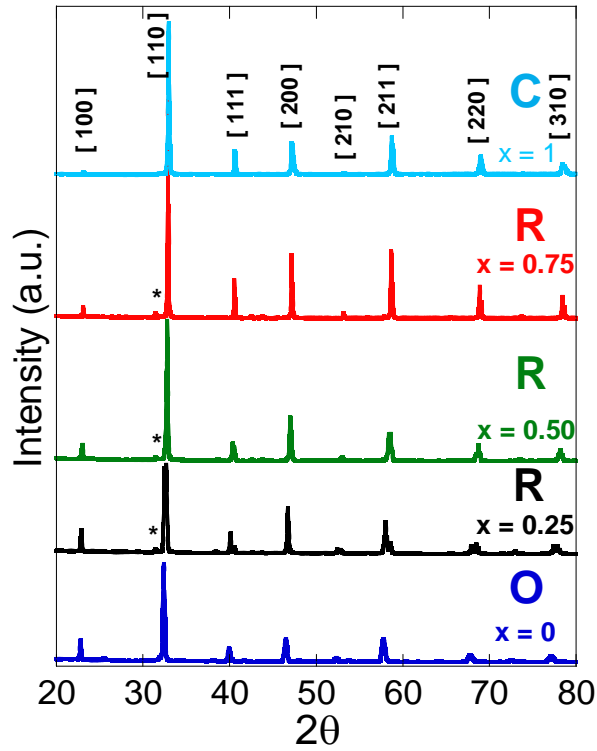

**Figure 1.** X-ray diffractograms for different compositions of the series $La_{1-x}Sr_xFe_{0.75}Ni_{0.25}O_{3-\delta}$. (x=0 orthorhombic; x=0.25, 0.5, 0.75 rhombohedral + little tetragonal; x=1 two cubic phases). The pure perovskite phase is indexed with all reflection peaks as the cubic phase with space group Pm-3m on top of the patterns. The extra peaks shown with the asterisks can possibly be assigned to the tetragonal phase $(La_{1.71}Sr_{0.19})NiO_{3.9}$.

**Electrical conductivity**

The electrical conductivity of the samples in the formula of $La_{1-x}Sr_xFe_{0.75}Ni_{0.25}O_{3-\delta}$ were measured in the temperature range of 300-1273 K, Fig. 2. All samples except for x= 0 show globally a similar temperature dependent behaviour: the conductivity increases up to a maximum value and then decreases. The transition temperature from polaronic or semiconducting conductivity behaviour to metallic like behaviour strongly depends on the relative Sr content. The sample for x=0.25 has the highest transition temperature (~ 1000 K). The sample with x=0.75 has two maxima at different temperatures, $T_1 \sim 985$ K and $T_2 \sim 734$ K. The effect of the temperature on the electrical conductivity of $La_{1-x}Sr_xFeO_3$ series is well studied and has been described in terms of small-polaron hopping conduction. The temperature dependent conductivity for small polaron conductors is given by an exponential of the form $\sigma \cdot T \sim \exp(E_p/kT)$ [12-17] where $E_P$ is the activation energy for small polaron hopping and k is the Boltzmann constant. The activation energies for our samples were determined from the Arrhenius plot of $\sigma T - 1/T$ for the low temperature region (305 K - 465 K) and high temperature region (655 K - 1230 K) and are listed in Table 2. The sample with x=0 shows the highest activation energy for both low and high temperature region. Once the La substituted by Sr, the electrical conductivity increases in contrast to that the activation energy decreases. The sample with x=0.50 shows the highest conductivity and has the lowest activation energy (39 meV) for all temperature region. The more Sr substitution cause to decrease the conductivity and increase the activation energy not only in LSF but also in LSFN system. LSFN has generally a higher electrical conductivity than LSF. In the $La_{1-x}Sr_xFeO_3$ series the sample with x=0.50 has highest conductivity (352 S/cm at 823 K). It has been shown that extensive sintering time at 1400°C resulted in decreased conductivity [18]. The maximum conductivity of $La_{0.5}Sr_{0.5}Fe_{0.8}Ni_{0.2}O_3$ is 700 S/cm, and that of $La_{0.5}Sr_{0.5}Fe_{0.4}Ni_{0.6}O_3$ is 350 S/cm [1], while our sample $La_{0.5}Sr_{0.5}Fe_{0.75}Ni_{0.25}O_{3-\delta}$ has 678 S/cm.

**Table 2.** Activation energies and the electrical conductivity at RT of the series $La_{1-x}Sr_xFe_{0.75}Ni_{0.25}O_{3-\delta}$

|  | Activation energy (meV) |  | Conductivity at RT (S/m) |
| --- | --- | --- | --- |
|  | 305 K-465 K | 655 K-1230 K |  |
| **x=0.00** | 170.60 | 180.50 | 106.19 |
| **x=0.25** | 104.97 | 105.09 | 5655.3 |
| **x=0.50** | 39.0 | 39.0 | 67412.0 |
| **x=0.75** | 72.07 | 69.36 | 5892.8 |
| **x=1.0** | ------------- | --------- | --------- |

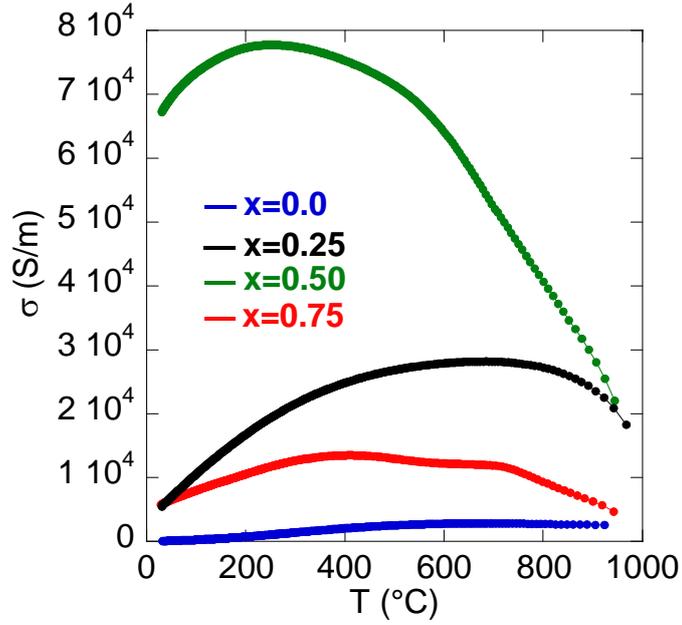

**Figure 2.** Temperature dependent electrical conductivity of $La_{1-x}Sr_xFe_{0.75}Ni_{0.25}O_{3-\delta}$ series.

**X-ray absorption spectra**

**I. Fe L2, 3 edge:**

The normalize Fe $L_{2,3}$ edge spectra of $La_{1-x}Sr_xFe_{0.75}Ni_{0.25}O_{3-\delta}$ series show the typical multiplets, $L_3$ ($2p_{3/2}$) and $L_2$ ($2p_{1/2}$) at about 705 eV and 718 eV due to the spin-orbit coupling of the Fe 2p core hole [8]. Each multiplet is additionally split by the crystal field effect into $t_{2g}$ and $e_g$ orbitals. The sample with x=0 has $Fe^{3+}$ configuration having a sharp $t_{2g}$ band in L3. Once La is substituted by Sr the spectra shift to higher energy, the intensity decreases and $t_{2g}$ in L3 edge becomes less sharp. Therefore, we can conclude that increasing the Sr content increases the oxidation state of Fe from 3+ to 4+. In order to improve our conclusion Atomic multiplet calculation [19] was made for all samples. The experimental spectra were at lower energy by ~8 eV and were shifted in order to get best agreement with the calculated spectra. Beside, the intensity of the experimental spectra was higher than the calculated one and was fixed to the same intensity of the calculated. Especially, the intensity and the energy position of the experimental spectra were tried to be fixed to the $e_g$ band in $L_3$. All the calculations were made in octahedral symmetry and Slater integrals scaled down to 50% of their atomic values in order to mimic covalency effects. Since it is not possible to calculate the linear combination of $Fe^{3+}$ and $Fe^{4+}$, we used another program to do it. All the comparisons between the experimental and theoretically calculated spectra are shown in Fig. 3. The comparison clearly shows that the sample with x=0 having sharp multiplet is in $3d^5$ ($Fe^{3+}$), in high spin state (S = 5/2) having $t_{2g}^3 e_g^2$ electronic configuration with ($^6A_{1g}$) symmetry and with crystal field 10Dq=1.85 eV, similar to the electronic configuration to LaFeO$_3$ [8]. The differences in the electronic structure of LaFeO$_3$ and LaFe$_{0.75}$Ni$_{0.25}$O$_3$ are Slater integrals and crystal field effect. Since our sample has lower Slater integrals, we can conclude that the Fe 3d orbitals becomes broader and d-d, p-d interactions and p-d exchange interactions decreases with adding 25% of Ni. Once Fe is substituted by Ni atoms, the M-O (M donates metals) distance is reduced causing an increase in the O 2p bandwidth. Thus, the p-d band gap becomes smaller and consequently the electrical resistivity decreases [20]. That is why LaFe$_{0.75}$Ni$_{0.25}$O$_3$ is a semiconductor while LaFeO$_3$ is an insulator at RT although they have the similar electronic configuration.

The comparison between experimental and theoretically calculated spectra shows that the sample with x=0.25 and x=0.5 have the same percentage of Fe$^{4+}$ with the Sr content. However, the sample with x=0.75 shows higher percentage of Fe$^{4+}$ than Sr content, 95%. All the samples are in high spin state with crystal field 10Dq=1.85 eV and with Slater integral reduced to 50%. Therefore, we can conclude that increasing the Sr content does not change the electronic structure parameters, only changes the oxidation state of Fe and Ni. The oxidation state of Fe which are calculated considering the Ni is 3+ and corrected by the experimental δ are quite close to that of obtained by atomic multiplet calculation except for high Sr doping level (x=0.75), Table 3.

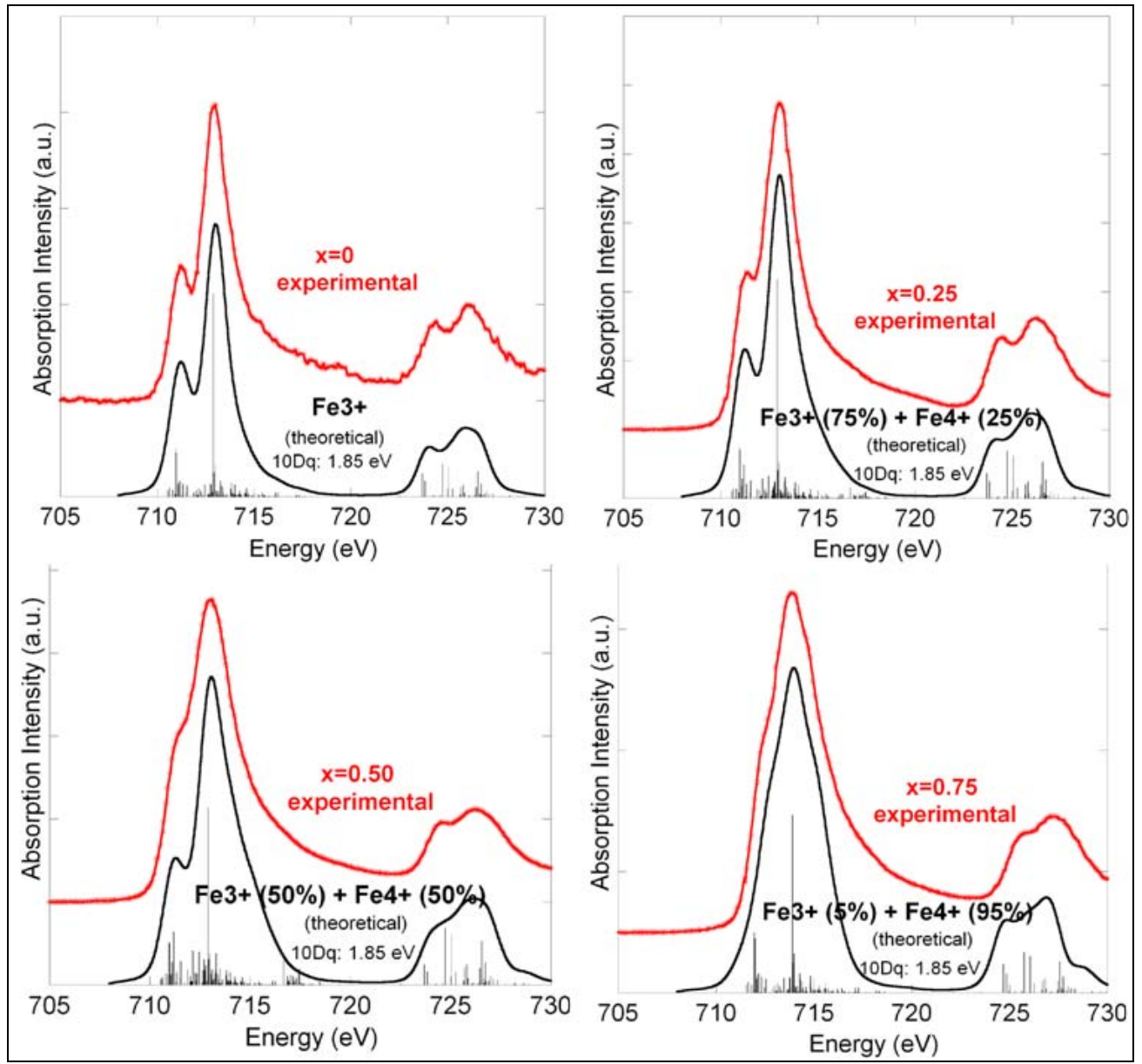

**Figure 3.** Comparison of experimental and theoretical x-ray absorption spectra of Fe L2, 3 edge.

## II. Oxygen K edge:

The Oxygen K edge (1s) XAS spectra in Fig. 4. represent the unoccupied oxygen 2p orbitals and also reflect the empty Fe/Ni 3d bands at the pre-edge (524 eV – 527 eV). The triplet at around 525 eV can be attributed to bands of $e_g(\uparrow)$, $t_{2g}(\downarrow)$ and $e_g(\downarrow)$ bands [21]. Comparing the pre-edge region of the spectra shown in Figure 6, we can easily identify that the intensity of $e_g(\uparrow)$ peak increases with increasing Sr content in the samples. High intensity in this region is due to doped holes, as we know from A-site substitution studies for LSF [8]. It is also important to impress that a pre-peak grows as one replaces $La^{3+}$ by $Sr^{2+}$ which means hole doping on oxygen site. The pre-peak shifts to Fermi level ($E_F$) depending on the hole/Sr doping concentration. The shifting is maximum for the sample with x=0.50 which has maximum conductivity. The first broad structure at higher energy (535 eV – 540) eV is attributed to La 5d and/or Sr 4d and the second broad structure starting at 540 eV is attributed to the bands of higher energy metal states Fe/Ni 4sp and La 6sp/ Sr 5sp.

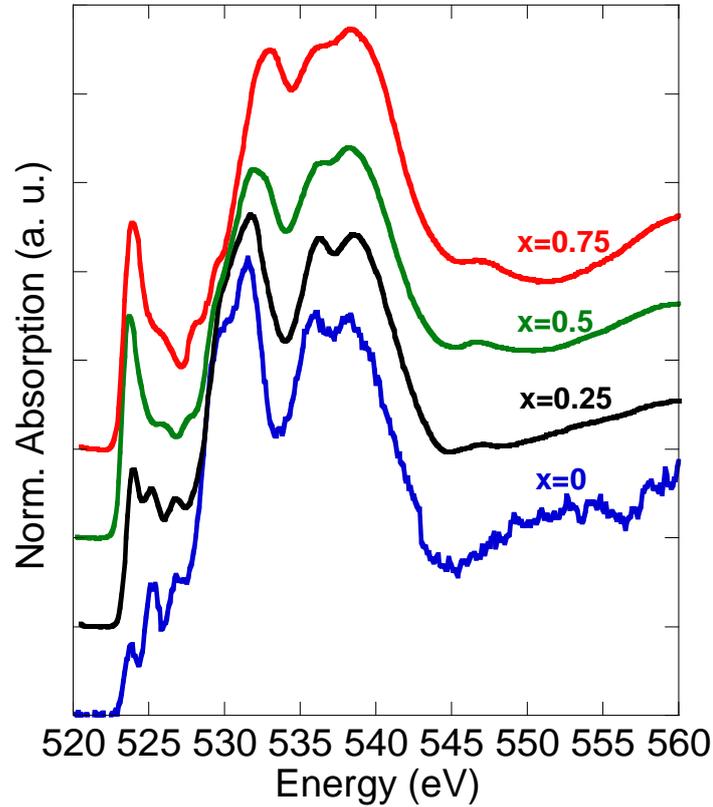

**Figure 4.** Normalized oxygen K edge x-ray absorption spectra of $La_{1-x}Sr_xFe_{0.75}Ni_{0.25}O_{3-\delta}$ series.

We want to illustrate the evolution of the electron hole concentration τ as a function of the A-site substitution, and begin with x=0 and with the simplified assumptions that δ=0, and that Ni has the oxidation state $Ni^{3+}$. Further, we may assume that La and Sr retain their formal valences as $La^{3+}$ and $Sr^{2+}$, i.e. a formal charge of -6 for $O_3$. The formal charge of the iron is thus $Fe^{+3.0}$ and the hole concentration τ = [$Fe^{4+}$] = 0. In LSF, the hole concentration τ relates to the stoichiometry as follows: $La_{1-x}Sr_xFe_{1-\tau}^{3+}Fe_\tau^{4+}O_{3-\delta}$. When we correct for the experimentally obtained δ for the other samples and listed them in Table 3. We have recently showed that the relative peak height S of spin up and spin down peaks of the upper Hubbard band in oxygen NEXAFS spectra (pre-edge peaks) correlates well with the conductivity in A- and B-site substituted LSF. We have exercised this as well for the oxygen spectra of our samples and found that the intensity ratio is not for all samples close to simple rational numbers, which we have also listed in Table 3. We can conclude that the p-projected electrons do not only goes to Fe 3d but also goes to Ni 3d.

**Table 3**: Oxygen non-stoichiometry δ and oxidation states for Fe, d-type hole concentrations τ and spectral ratio of $(e_g(\uparrow)/t_{2g}(\downarrow) + e_g(\downarrow))$ S calculated from oxygen K edge spectra in $La_{1-x}Sr_xFe_{0.75}Ni_{0.25}O_{3-\delta}$. The τ calculation is based on an oxidation state of 3+ for Ni.

| Sr content x | δ | Ox Fe for δ=0 | Ox Fe for δ=exp | Ox Fe from simulation | τ for δ=0 | τ for δ=exp | S |
|---|---|---|---|---|---|---|---|
| 0 | 0.000 | 3 0/3 (3.0) | 3.00 | 3.00 | 0 | 0 | 1/4 |
| 1/4 | 0.035 | 3 1/3 (3.33) | 3.24 | 3.25 | 1/2 | 0.316 | 1/3 |
| 1/2 | 0.053 | 3 2/3 (3.67) | 3.53 | 3.50 | 2 | 1.128 | 1/2 |
| 3/4 | 0.069 | 3 3/3 (4.0) | 3.82 | 3.95 | >>1 | 4.55 | 3/4 |

The evolution of the spectral ratio S of the oxygen pre-edge peaks which is linear in the substitution parameter x, and the relative hole density of Fe τ are shown in Fig. 5.

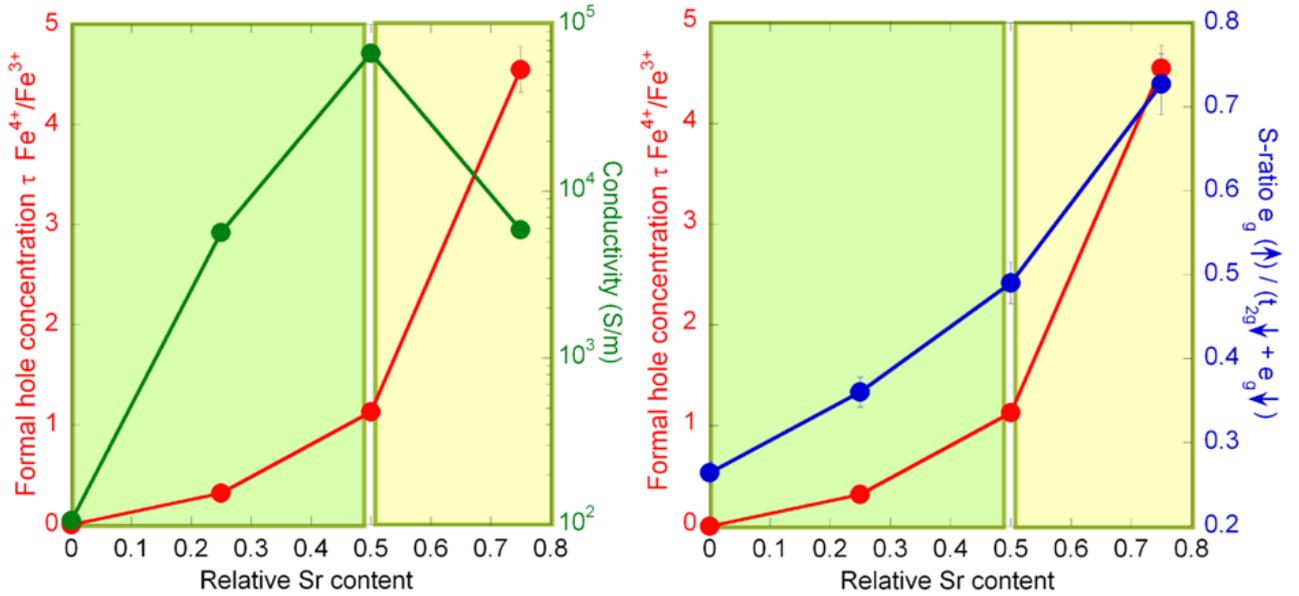

**Figure 5.** Comparison of d-type hole with conductivity ((left), and with spectral ratio of $(e_g(\uparrow)/t_{2g}(\downarrow) + e_g(\downarrow))$ calculated from the oxygen K edge spectra depending on x in $La_{1-x}Sr_xFe_{0.75}Ni_{0.25}O_{3-\delta}$.

Once $La^{3+}$ is substituted by $Sr^{2+}$, one hole is created. The hole concentration τ ($Fe^{4+}/Fe^{3+}$) of LSF-Ni increases with increasing Sr content, Fig. 5. The conductivity and τ increases with increasing Sr content up to 50%. In high doping region, the conductivity decreases while τ increases. It is known for LSF that in low Sr doping region the mobility of free electrons increases with increasing Sr content: coulomb repulsions induced by $La^{3+}$ ions around oxygen anion in the chain ($Fe^{4+}$-$O^{2-}$-$Fe^{3+}$) stronger than the barrier created by $Sr^{2+}$ having smaller charge. In high Sr doping region, oxygen vacancies along the migration pathway increase [22]. Therefore, for an electron it becomes more difficult (needs more energy) to jump from ($Fe^{3+}$-$O^{2-}$-$Fe^{4+}$) which results in a decrease in mobility of electron and/or holes consequently decrease in electrical conductivity. This conclusion is also proved by an increase in hopping activation energy, Ea=72.07 meV for x=0.75 while Ea=0.39 meV for x=0.50. 50% doping on the A-site where $\tau = 1$ is also known for $La_{1-x}Sr_xFeO_3$ [8]. The magnetic interation of $Fe^{3+}$-O-$Fe^{3+}$ is antiferromagnetic while the $Fe^{3+}$-O-$Fe^{4+}$ is ferromagnetic [26]. Once $Fe^{3+}$ is oxidized to $Fe^{4+}$, the number of the unpaired electrons decreases and antiferromagnetic structure is destroyed due to the ferromagnetic interaction. It is also known for $Sr_xHo_{1-x}FeO_{3-\delta}$ that the conductivity increases with increasing $\tau$ [23].

We compare the p-type and d-type hole concentration depending on relative Sr content, in Fig. 5. P-type holes are created on oxygen site by transition of electron from oxygen 2p to Fe/Ni 3d which is proportional to the intensity of the ratio of $e_g(\uparrow)/t_{2g}(\downarrow) + e_g(\downarrow)$. If only we have one metal on the B-site, it is possible to conclude that the electrons transfer to that B-site metal with creating hole on oxygen site. However, here we have 2 different metals on B-site, Fe and Ni. Therefore, only looking at the p-type hole concentration it is not possible to distinguish whether the electrons goes to Fe or Ni 3d orbitals. We show the d-type hole for Fe but not for the Ni due to the hybridization between La and Ni 3d orbitals we were not able to get pure Ni absorption spectra. Starting from $LaFe_{0.75}Ni_{0.25}O_3$ which is in $3d^5$ configuration and not in $3d^5\underline{L}$ which means there is no charge transfer from oxygen to Fe. However, in the oxygen spectra there is a pre-peak with the ratio of $e_g(\uparrow)/t_{2g}(\downarrow) + e_g(\downarrow)$ =0.25. We can conclude that this charge transfer occurred from oxygen 2p to Ni 3d. It is known that Fe ($\varepsilon_d$-$\varepsilon_p$) > Ni ($\varepsilon_d$-$\varepsilon_p$) where $\varepsilon_d$ and $\varepsilon_p$ are the energy of 3d states and 2p states of Fe/Ni and oxygen, respectively [24]. As it is shown in Fig. 5 (right) increasing the Sr content increases both oxidation state of Fe and S-ratio. If the p-electrons go to Fe 3d orbitals Fe should become reduced. Therefore, we can also conclude for the Sr doped samples the charge transfer occurred mostly between oxygen 2p and Ni 3d.

## CONCLUSIONS

In $La_{1-x}Sr_xFe_{0.75}Ni_{0.25}O_{3-\delta}$ we showed that increasing the Sr content increases both d-type ($Fe^{4+}/Fe^{3+}$) and p-type holes [($e_g(\uparrow)/t_{2g}(\downarrow) + e_g(\downarrow)$)]. We conclude that the charge transfer occurred mostly from oxygen to Ni and not from oxygen to Fe. The electrical conductivity which has contribution from both side of d- and p- holes increases up to 50% of Sr doping because once $La^{3+}$ is replaced by $Sr^{2+}$ the coulomb interaction is reduced consequently potential barrier is reduced. Thus, electrons or electron holes can easily move from one site to another ($Fe^{3+}$-$O^{2-}$-$Fe^{4+}$). In low Sr doping level, we found that the percentage of $Fe^{4+}$ determined by atomic multiplet calculation is similar to the percentage of Sr content. In high Sr doping region, the electrical conductivity starts to decrease while the both d- and p-type hole concentration increases because of the vacancies created along the migration pathway. The phase transformation from orthorhombic (x=0.0) to rhombohedral (x=0.25, 0.50, 0.75) and finally cubic (x=1) is observed. Increasing the Sr content increases the percentage of $Fe^{4+}$ configuration and cause but does not affect the spin state of Fe.

**Acknowledgment**

Financial support by Swiss National Science Foundation, project # NSF No.200021-116688 and by the European Commission, contract # MIRG No. CT-2006-042095 is gratefully acknowledged. The ALS is supported by the Director Office of Science/BES, of the U. S. DoE, No. DE-AC0205CH11231.

**Figure captions:**

Figure 1. X-ray diffractograms for different compositions of the series $La_{1-x}Sr_xFe_{0.75}Ni_{0.25}O_{3-\delta}$. (x=0 orthorhombic; x=0.25, 0.5, 0.75 rhombohedral + little tetragonal; x=1 two cubic phases). The pure perovskite phase is indexed with all reflection peaks as the cubic phase with space group Pm-3m on top of the patterns. The extra peaks shown with the asterisks can possibly be assigned to the tetragonal phase $(La_{1.71}Sr_{0.19})NiO_{3.9}$.

Figure 2. Temperature dependent electrical conductivity of $La_{1-x}Sr_xFe_{0.75}Ni_{0.25}O_{3-\delta}$ series.

Figure 3. Comparison of experimental and theoretical x-ray absorption spectra of Fe L2, 3 edge.

Figure 4. Normalized oxygen K edge x-ray absorption spectra of the $La_{1-x}Sr_xFe_{0.75}Ni_{0.25}O_{3-\delta}$ series.

Figure 5. Comparison of d-type hole with conductivity ((left), and with spectral ratio of ($e_g(\uparrow)/t_{2g}(\downarrow) + e_g(\downarrow)$)) calculated from the oxygen K edge spectra depending on x in $La_{1-x}Sr_xFe_{0.75}Ni_{0.25}O_{3-\delta}$.

**Table captions:**

Table 1. Refined structural parameters for the $La_{1-x}Sr_xFe_{0.75}Ni_{0.25}O_{3-\delta}$ series.

Atomic positions for Pbnm are 4(c) (0, 0, 1/4) for La; 4(b) (0, 0, 1/4) for Fe/Ni and (4c) (1/2, 1/4, 0) for O1 and (8d) (1/4, 1/2, 1/4) for O2. Atomic positions for R-3c are 6(a) (0, 0, 1/4) for La; 6(b) (0, 0, 0) for Fe/Ni and 18(e) (1/2, 0, 1/4) for O. Atomic positions for Pm-3m are 6(a) (1/2, 1/2, 1/2) for Sr; 6(b) (0, 0, 0) for Fe/Ni and 18(e) (1/2, 0, 0) for O.

Table 2. Activation energies and the electrical conductivity at RT of the series $La_{1-x}Sr_xFe_{0.75}Ni_{0.25}O_{3-\delta}$

Table 3. Oxygen non-stoichiometry δ and oxidation states for Fe, d-type hole concentrations τ and spectral ratio of ($e_g(\uparrow)/t_{2g}(\downarrow) + e_g(\downarrow)$)  S calculated from oxygen K edge spectra in $La_{1-x}Sr_xFe_{0.75}Ni_{0.25}O_{3-\delta}$. The τ calculation is based on an oxidation state of 3+ for Ni.